\documentclass[prl,twocolumn,reprint,amsmath,amssymb,superscriptaddress,floatfix]{revtex4}
\usepackage[breaklinks=true,colorlinks,citecolor=blue,linkcolor=blue,urlcolor=blue]{hyperref}
\usepackage{float}
\usepackage{textcomp}
\usepackage{epsfig,mathrsfs,color,latexsym,subfigure,marginnote}

\def\be{\begin{equation}}
\def\ee{\end{equation}}

\def\bea{\begin{eqnarray}}
\def\eea{\end{eqnarray}}

\begin{document}

\title{Topological Hourglass Dirac Semimetal in the Nonpolar Phase of Ag$_2$BiO$_3$ }

\author{Bahadur Singh\footnote{These authors contributed equally to this work.}$^{\dag}$}
\affiliation{SZU-NUS Collaborative Center and International Collaborative Laboratory of 2D Materials for Optoelectronic Science $\&$ Technology, Engineering Technology Research Center for 2D Materials Information Functional Devices and Systems of Guangdong Province, College of Optoelectronic Engineering, Shenzhen University, ShenZhen 518060, China}
\affiliation{Department of Physics, Northeastern University, Boston, Massachusetts 02115, USA}

\author{Barun Ghosh$^*$}
\affiliation{Department of Physics, Indian Institute of Technology Kanpur, Kanpur 208016, India}

\author{Chenliang Su$^{\dag}$}
\affiliation{SZU-NUS Collaborative Center and International Collaborative Laboratory of 2D Materials for Optoelectronic Science $\&$ Technology, Engineering Technology Research Center for 2D Materials Information Functional Devices and Systems of Guangdong Province, College of Optoelectronic Engineering, Shenzhen University, ShenZhen 518060, China}

\author{Hsin Lin$^{\dag}$}
\affiliation{Institute of Physics, Academia Sinica, Taipei 11529, Taiwan}

\author{Amit Agarwal\footnote{Corresponding authors' emails: 
bahadursingh24@gmail.com, chmsuc@szu.edu.cn, nilnish@gmail.com, amitag@iitk.ac.in}}
\affiliation{Department of Physics, Indian Institute of Technology - Kanpur, Kanpur 208016, India}

\author{Arun Bansil}
\affiliation{Department of Physics, Northeastern University, Boston, Massachusetts 02115, USA}


\begin{abstract}
Materials with tunable charge and lattice degrees of freedom provide excellent platforms for investigating multiple phases that can be controlled via external stimuli.  We show how the charge-ordered ferroelectric oxide Ag$_2$BiO$_3$, which has been realized experimentally, presents a unique exemplar of a metal-insulator transition under an external electric field. Our first-principles calculations, combined with a symmetry analysis, reveal the presence of a nearly ideal hourglass-Dirac-semimetal state in the nonpolar structure of Ag$_2$BiO$_3$. The low-energy band structure consists of two hourglasslike nodal lines located on two mutually orthogonal glide-mirror planes in the absence of spin-orbit coupling (SOC) effects. These lines cross at a common point and form an interlinked chainlike structure, which extends beyond the first Brillouin zone. Inclusion of the SOC opens a small gap in the nodal lines and results in two symmetry-enforced hourglasslike Dirac points on the $\tilde{\mathcal{C}}_{2y}$ screw rotation axis. Our results indicate that Ag$_2$BiO$_3$ will provide an ideal platform for exploring the ferroelectric-semiconductor to Dirac-semimetal transition by the application of an external electric field.

\end{abstract}

\maketitle
Symmetry-protected topological states of matter have been at the forefront of quantum materials research over the last decade \cite{Bansil16,Hasan10,Qi11,Fu07,Fu_TCI}. By considering various crystalline symmetries, a variety of topological semimetals with their own unique properties have been proposed, and a number of these proposals have been realized experimentally \cite{Bansil16,Hasan10,Qi11,Fu07,Fu_TCI,Wan11,Singh12,Burkov14,Kane_DSM_in3d,Yang2014,cubic_dirac,Liang2015,Fu_nodallineSM,Bian2016,Bradlyn,Po2017}.  Well-known examples include Dirac and Weyl semimetals, which have been realized in Na$_3$Bi, Cd$_3$As$_2$, TaAs, LaAlGe, and MoTe$_2$, among other materials \cite{Wang13b,Liu864,Wang13a,Liu2014,Huang15,Yang2015,Xu613,TaAs_expt,Xue1603266,Deng2016, Jiang2017}. Recently, it has been recognized that nonsymmorphic crystal symmetries can provide a new mechanism for stabilizing band-crossing points in materials \cite{Kane_DSM_in2d,Zhao16,Yu17,Bzdusek2016}. In this way, multiple bands can become entangled and their crossing points can become inevitable and entirely mandated by crystalline symmetries. Such essential band crossings have been predicted on the 2D surface of nonsymmorphic insulators KHgX ({\it X }= As, Sb, Bi) where four bands mix together to form an hourglasslike energy dispersion \cite{Wang2016,Ma2017}. Other theoretical proposals have shown that the hourglass fermions can exist in 3D bulk bands, and that even a nodal-chain structure can be generated in the presence of multiple nonsymmorphic symmetries \cite{Takahasi17,Wang17,Wang2017,Li18}.  

\begin{figure}
\includegraphics[width=0.5\textwidth]{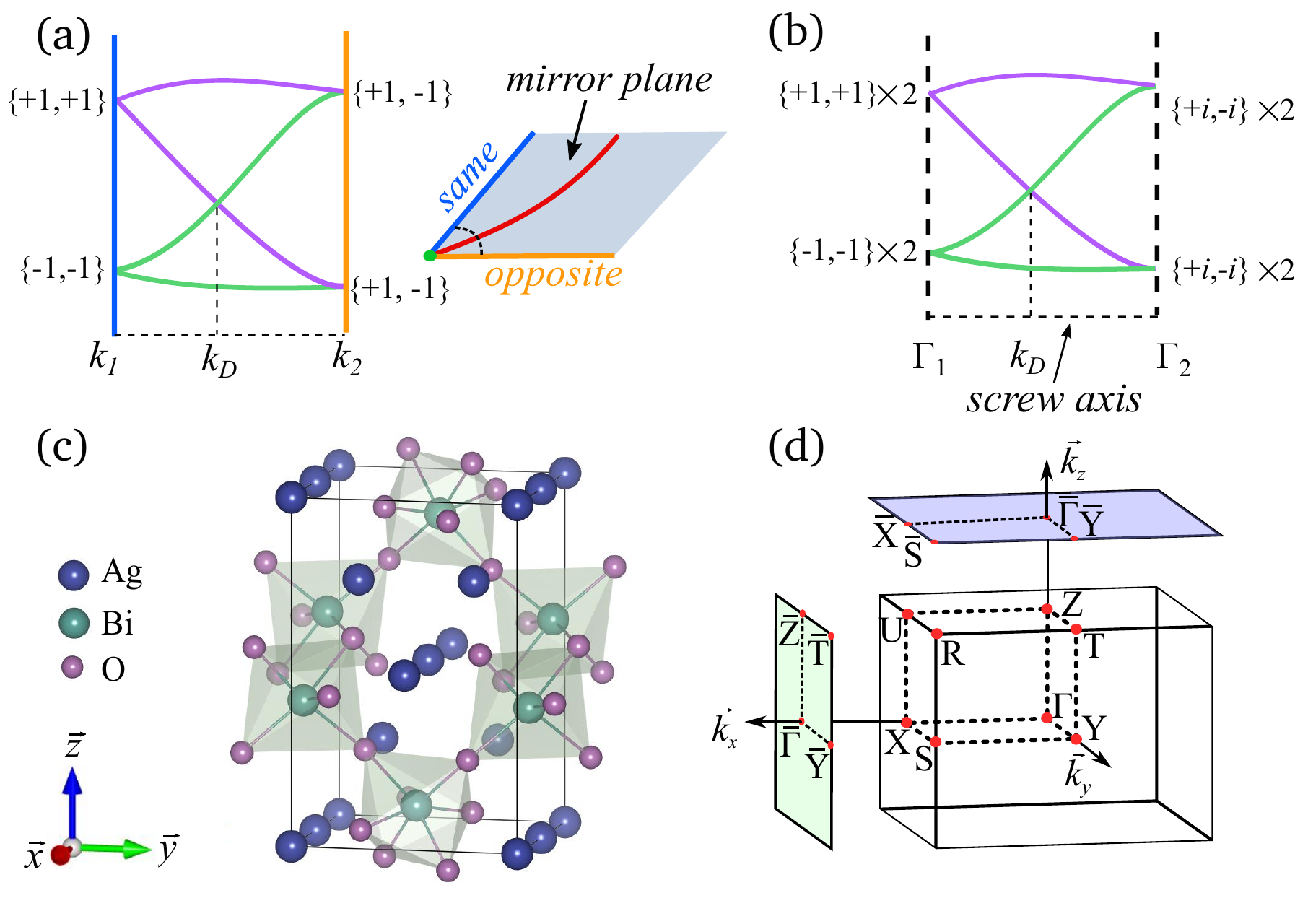}
\caption{Schematic of the hourglasslike energy dispersion (a) on a glide mirror plane between two high-symmetry lines (without SOC), and (b) on a screw rotation axis between two time-reversal invariant momentum points (with SOC). The mirror eigenvalues in (a) and the rotation eigenvalues in (b) are shown. Hourglass band crossings trace a protected nodal line on the mirror plane and form a Dirac point on the screw rotation axis. (a) and (b) represent Ag$_2$BiO$_3$ without and with SOC, respectively (see text for details). (c) Crystal structure of nonpolar Ag$_2$BiO$_3$. (d) Bulk and projected (001) and (100) surface BZs. Various high symmetry points are marked.}
\label{fig:CS}
\end{figure}

Figures \ref{fig:CS}(a)-\ref{fig:CS}(b) schematically show the glide-mirror and screw-rotation symmetry-driven hourglass semimetal energy dispersions. Even though the hourglass fermions are dictated by space group symmetries, establishing their existence in the experimentally synthesized materials has proven challenging. This is due in part to the presence of trivial band features around the Fermi energy in the proposed materials that harbor hourglasslike dispersions. Also, in 3D materials with nodal lines, identification of topological features becomes uncertain due to the presence of the background of many other bands with large dispersions. 

In this Letter, based on our first-principles computations, we predict the presence of a nearly ideal hourglass-fermion state in the nonpolar phase of Ag$_2$BiO$_3$, a material which has been shown to be a charge-ordered ferroelectric semiconductor at room temperature. Ferroelectricity is an intriguing phenomenon that is driven by the formation of electric dipoles, which arise when the centers of positive and negative charges are separated under a polar distortion. Application of an external electric field in the direction opposite to that of the spontaneous polarization can be used to attenuate the distortion \cite{ferroelectric_metal,ferroE_switch,He2018}. Ag$_2$BiO$_3$ has been shown to exist in three distinct crystal structures of nonpolar ({\it Pnna}) and polar ({\it Pnn}2 and { \it Pn}) space groups. The nonpolar and polar structures are paraelectric semimetal and ferroelectric semiconductor, respectively, where the polar structure contains a centrosymmetric BiO$_6$ octahedral breathing distortion \cite{He2018,DEIBELE1999117,OBERNDORFER2006267}. This distortion has been attributed to the coexistence of two valence states, Bi$^{3+}$ and Bi$^{5+}$, due to charge disproportionation of the formal Bi valence state, which induces a sizable ferroelectric polarization. An external electric field can then nullify charge disproportionation and transition the system into the metallic nonpolar {\it Pnna} structure. Such a transition has also been observed in other Bi oxides \cite{BaBiO3,BaBiO3_2,Shi2013}. 

Our analysis here reveals that nonpolar {\it Pnna} Ag$_2$BiO$_3$ supports an exotic hourglass nodal-chain-semimetal state when the spin-orbit coupling (SOC) is neglected. Inclusion of the SOC gaps out the nodal lines and results in the appearance of additional symmetry-enforced Dirac cones on the screw rotation axis. In order to showcase the nontrivial topology of the bulk bands, we have computed the topological surface states supported by various crystal surfaces. In this way, we show that Ag$_2$BiO$_3$ is a unique material that undergoes ferroelectric semiconductor to hourglass semimetal transition under the application of an external electric field. 

Electronic structure computations were carried out within the framework of the density functional theory (DFT) with the projector-augmented-wave (PAW) pseudopotentials using the VASP code \cite{Kohn_dft,vasp}. Exchange-correlation effects were treated at the level of the generalized gradient approximation with PBE parametrization \cite{Perdew_gga}. We used an energy cutoff of 500 eV for the plane wave basis set and a $k$ mesh of $13\times12\times8$ for the Brillouin zone (BZ) sampling. Starting with the experimental crystal structure, we optimized both the atomic positions and lattice parameters. The tight-binding parametrization used was obtained through the VASP2WANNIER90 interface and the surface spectrum was obtained by using the iterative Green's function approach as implemented in the WannierTools package \cite{wannier,Wu2017}.

Based on a representation theory analysis, the nonpolar {\it Pnna} group (No. 52) is a supergroup of the polar {\it Pnn}2 group (No. 34) \cite{Bilbao}. The nonpolar {\it Pnna} structure can thus be considered as the high-symmetry parent phase of the polar {\it Pnn}2 structure. The crystal lattice of nonpolar Ag$_2$BiO$_3$ is shown in Fig.~\ref{fig:CS}(c). The associated bulk BZ and its projections on the (100) and (001) planes are shown in Fig.~\ref{fig:CS}(d). There are 24 atoms in the unit cell where the  four Bi atoms form an octahedral coordination structure with the O atoms while Ag atoms fill the interstitial sites. The Bi-O bond length lies between the ideal Bi$^{\rm{ +3} }$-O and Bi$^{\rm {+5}}$-O bond lengths. This crystal hosts many symmetries, which include the screw rotation 
$\tilde{\mathcal{C}}_{2y}: (x,y,z)\rightarrow(-x+\frac{1}{2},y+\frac{1}{2},-z+\frac{1}{2})$, and three glide  mirror planes
$\tilde{\mathcal{M}}_{x}: (x,y,z)\rightarrow(-x,y+\frac{1}{2},z+\frac{1}{2})$,
$\tilde{\mathcal{M}}_{y}: (x,y,z)\rightarrow(x+\frac{1}{2},-y+\frac{1}{2},z+\frac{1}{2})$, and
$\tilde{\mathcal{M}}_{z}: (x,y,z)\rightarrow(x+\frac{1}{2},y,-z)$. Furthermore, the crystal respects inversion ($\mathcal{I}$) and time-reversal ($\Theta$) symmetries.

\begin{figure}[t]
\includegraphics[width=0.49\textwidth]{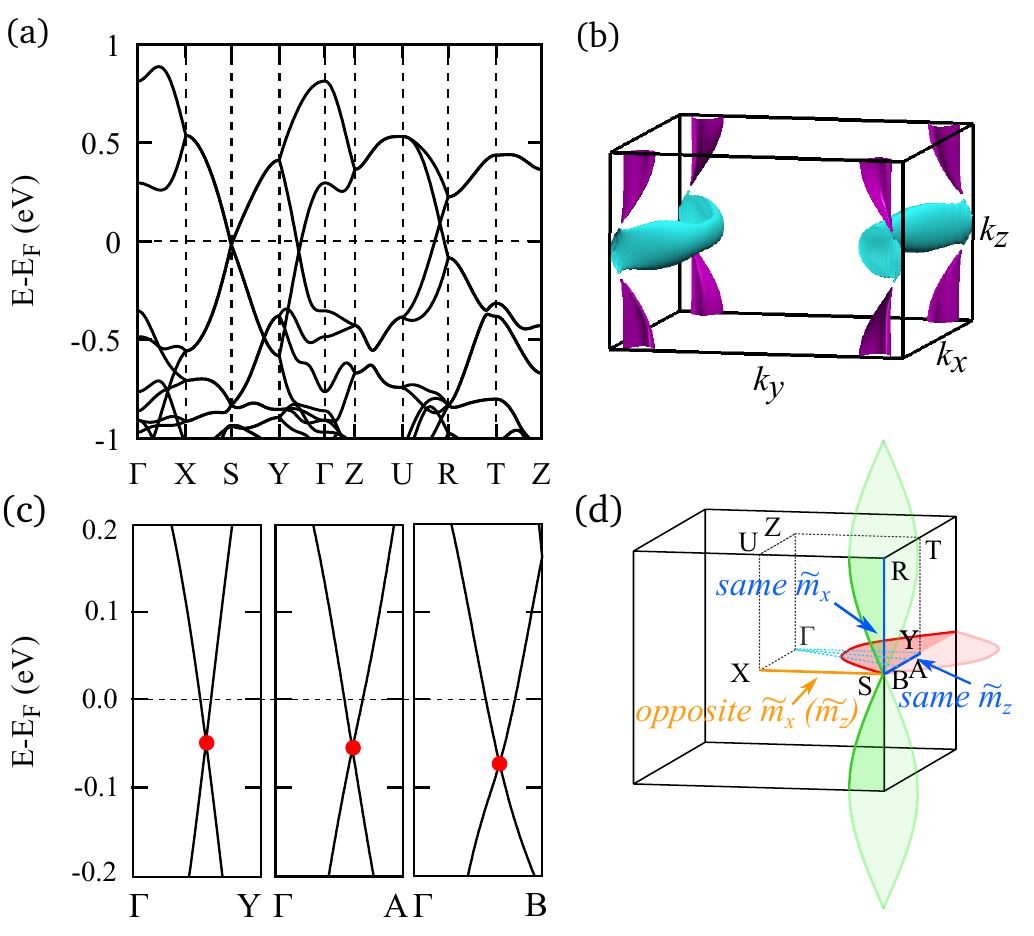}
\caption{(a) Band structure of {\it Pnna} Ag$_2$BiO$_3$ (without SOC). The linear band crossings are evident near the Fermi level (horizontal dashed line). (b) The Fermi surface depicting the electron (cyan) and hole (violet) pockets. (c) Band structure on the $k_z=0$ plane along selected directions, which are marked by cyan lines in (d). Band crossings are linear and show very small energy spread. (d) The momentum space distribution of hourglass nodal lines in the BZ. Red and green colors identify nodal lines on the $k_z=0$ and $k_x=\pi$ plane, respectively. Blue and orange thick lines show the high-symmetry directions with double band degeneracy of same and opposite mirror eigenvalues, respectively.}\label{fig:nosoc}
\end{figure}

The band structure of nonpolar Ag$_2$BiO$_3$ (without SOC) is semimetallic, see Fig.~\ref{fig:nosoc}(a). The valence and conduction bands are seen to cross linearly along the high-symmetry lines $\Gamma-Y$ and $U-R$ as well as the high-symmetry point $S$. There are no other (trivial) band crossings in an energy window $\pm 300$ meV around the Fermi level, and the bands remain at least twofold degenerate (excluding physical spin) at the high-symmetry lines on the BZ boundary. A careful inspection of band crossings in the full BZ reveals that the crossing points trace nodal lines on the $k_z=0$ and $k_x= \pi$ planes. These nodal lines intersect each other at $S$, forming an interlinked nodal-chain structure that extends beyond the first BZ, see Fig.~\ref{fig:nosoc}(d). The nodal point $S$ of the nodal chain is a fourfold degenerate Dirac point with linear energy dispersion along all three principal momentum directions. Note that the formation of nodal lines involves four bands around the Fermi level, which is a clear signature of an hourglass semimetal. We have further explored the spread of nodal lines in energy by plotting the band structure along various line cuts in the $k_z=0$ plane in Fig. \ref{fig:nosoc}(c). The band crossings are seen to trace a torus-shaped contour near the Fermi level with nearly linear energy dispersion. The Fermi surface in Fig. \ref{fig:nosoc}(b) shows two torus-shaped electron (cyan) and hole (violet) pockets, which originate from the nodal lines on the $k_z=0$ and $k_x=\pi$ planes, respectively.  In particular, the $k_z=0$ nodal line yields the electron pockets whereas the $k_x=\pi$ nodal line forms the hole pockets. These unique spectral features of Ag$_2$BiO$_3$ could also result in unique transport properties in this material.      
 
In order to understand the origin and stability of hourglass nodal lines, we start by examining the nature of the double band degeneracy along the high symmetry $XS$, $YS$ and $RS$ directions. As discussed in Ref.~[\onlinecite{Takahasi17}], this degeneracy in a nonsymmorphic spinless system arises from the antiunitary operator $\tilde{\mathcal{G}}\Theta$, which is formed by combining time-reversal symmetry $\Theta$ and the glide operator $\mathcal{\tilde{G}}$. Let us first consider the $XS$ line, which is an invariant subspace of the glide operators $\tilde{\mathcal{M}}_x$ and $\tilde{\mathcal{M}}_z$ with $\tilde{\mathcal{M}}_x^2=e^{-ik_y}$ and $\tilde{\mathcal{M}}_z^2=-1$. The associated glide eigenvalues are $\tilde{m}_x=\pm e^{-ik_y/2}$ and $\tilde{m}_z=\pm i$. The corresponding antiunitary operator that commutes with the Hamiltonian on this line is $\tilde{\mathcal{M}_y}\Theta$ since $(\tilde{\mathcal{M}}_y\Theta)^2=e^{-ik_x-ik_z}=-1$. This results in the double degeneracy of states $|u\rangle$ and $\tilde {\mathcal{M}}_y\Theta |u\rangle$ on the $XS$ line. It can be easily shown that these degenerate states have opposite $\tilde{\mathcal{M}}_x$ and $\tilde {\mathcal{M}}_z$ eigenvalues. Consider, for example, a Bloch state $|u\rangle$ with the positive $\tilde{\mathcal{M}}_x$ eigenvalue $ +e^{-ik_y/2}$, {\it i.e} $\tilde{\mathcal{M}}_x|u\rangle= + e^{-ik_y/2} |u\rangle$. 
Recalling the relationship $\tilde{\mathcal{M}}_x\tilde{\mathcal{M}}_y\Theta=-e^{-ik_y}\tilde{\mathcal{M}}_y\Theta\tilde{\mathcal{M}}_x$, the eigenvalue of its degenerate partner $\tilde{\mathcal{M}}_y\Theta |u\rangle$ is $-e^{-ik_y/2}$. A similar analysis can be used to show (i) the antiunitary operators $\tilde{\mathcal{M}}_x\Theta$, $\tilde{\mathcal{M}}_z\Theta$, and $\tilde{\mathcal{C}}_{2y}\Theta$ dictate the double degeneracy along the $YS$, $RS$, and $k_y=\pi$ plane, respectively, and, (ii) the degenerate bands along these paths have same eigenvalues for the operators $\tilde{\mathcal{M}}_x$ and $\tilde{\mathcal{M}}_z$. The $\tilde{\mathcal{M}}_x$ and $\tilde{\mathcal{M}}_z$ eigenvalues of the degenerate states on the high-symmetry lines are marked in Fig. \ref{fig:nosoc}(d).

We now turn to discuss the hourglass crossings in Ag$_2$BiO$_3$. These crossings appear on the intermediate $k$ points connecting two high-symmetry lines where the degenerate bands possess the same and opposite glide-mirror eigenvalues. The partner switching between these two lines is mandated by the nonsymmorphic symmetry, see Fig.~\ref{fig:CS}(a) for an illustration. From the symmetry analysis presented above, it is clear that the symmetry lines with degenerate partners of same and opposite eigenvalues of $\tilde{\mathcal{M}}_x$ and $\tilde{\mathcal{M}}_z$ exist on $k_x=\pi$ and $k_z=0$ planes (Fig.~\ref{fig:nosoc}(d)). Therefore, $k_x=\pi$ and $k_z=0$ planes must have intermediate $k$ points lying between the two lines with hourglass nodal band crossings, which are shown in Fig.~\ref{fig:nosoc}(d). In any event, independent to the symmetry analysis, we have confirmed the existence of the aforementioned nodal lines and the associated partner switching by explicitly calculating the  $\tilde{\mathcal{M}}_z$ and  $\tilde{\mathcal{M}}_x$ eigenvalues on $k_z=0$ and $k_x=\pi$ planes, respectively. The topological protection of nodal lines was further checked by calculating the nontrivial $\pi$ Berry phase on a small loop interlinking the nodal line crossing.

\begin{figure}
\includegraphics[width=0.49\textwidth]{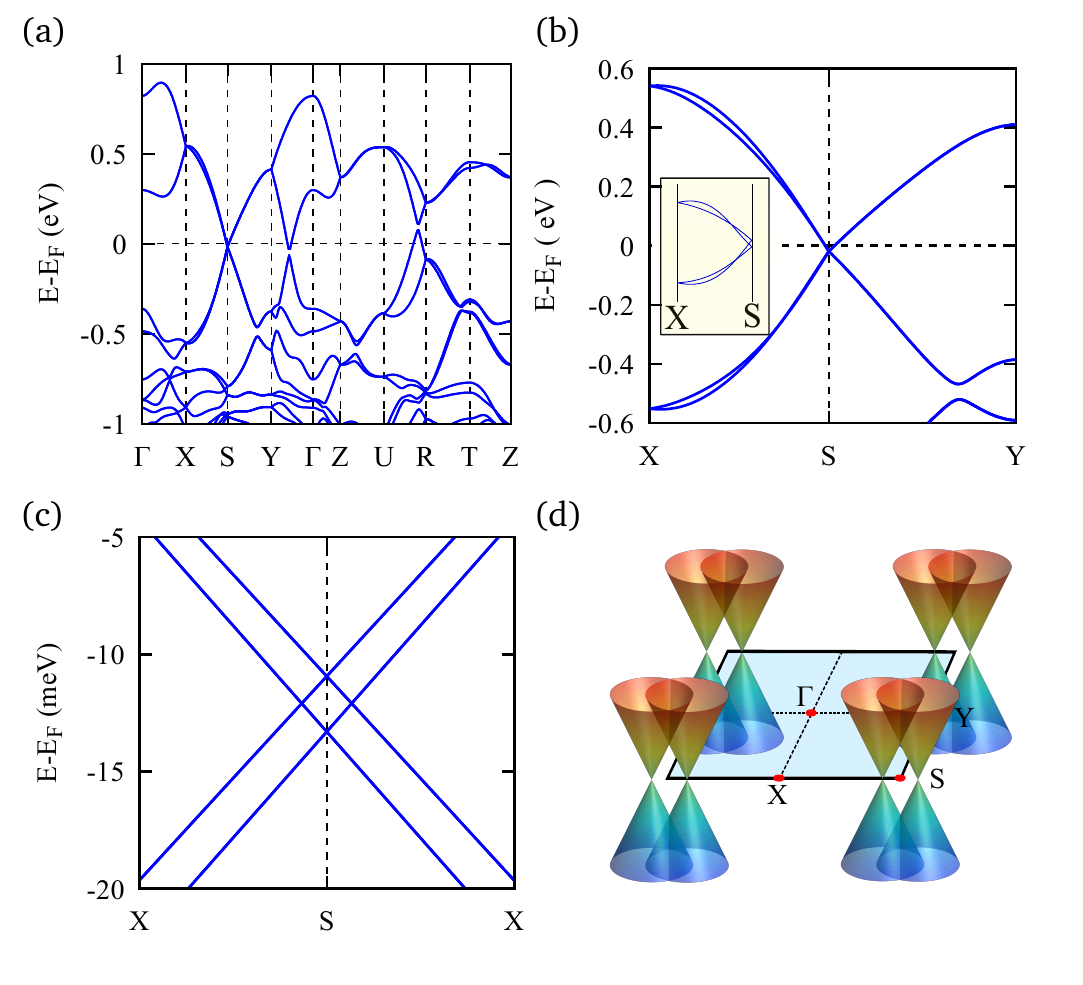}
\caption{(a) Band structure of {\it Pnna} Ag$_2$BiO$_3$  with SOC. (b) Closeup of the energy dispersion along the $XSY$ direction. Inset schematically shows the bands along the $XS$ line. (c) Energy dispersion in the vicinity of the $S$ point. Two hourglass Dirac points are seen to appear along the $XS$ direction. (d) Schematic of the hourglass Dirac semimetal state in Ag$_2$BiO$_3$ with four Dirac cones on the $k_z=0$ plane in the first BZ. }
\label{fig:wsoc}
\end{figure} 
In Fig. \ref{fig:wsoc}, we present the energy bands with SOC and explore the emergence of the hourglass Dirac fermions in Ag$_2$BiO$_3$. Owing to the presence of both $\mathcal{I}$ and $\Theta$ in the crystal, each band possesses an additional twofold Kramers's degeneracy. Also, the glide-mirror and screw-rotation operators are modified due to the inclusion of spin degrees of freedom. As a result, the nodal lines are now gapped with the appearance of a small $k$-dependent band gap at the band crossing points, see Fig. \ref{fig:wsoc}(a). Interestingly, despite the presence of the gapped nodal-line-crossings (with SOC), we find that the band structure still remains gapless. There are two additional four-fold degenerate Dirac points along the $XS$ high-symmetry line as shown in Figs.~\ref{fig:wsoc}(b)-\ref{fig:wsoc}(d). These results show that Ag$_2$BiO$_3$ undergoes a topological phase transition from being an hourglass-nodal-line semimetal to the hourglass-Dirac-semimetal state.

The hourglass Dirac points in the presence of the SOC are protected by $\tilde{\mathcal{C}}_{2y}$ screw-rotation symmetry. Since the $XS$ line is an invariant subspace of the $\tilde{\mathcal{C}}_{2y}$ operation, Bloch states on this line can have definite $\tilde{\mathcal{C}}_{2y}$ eigenvalues. $\tilde{\mathcal{C}}_{2y}^2=-e^{-ik_y}$ then leads to eigenvalues of the operator $\tilde{\mathcal{C}}_{2y}$ of $\pm i e^{-ik_y/2}$. Specifically, at the $X$ ($\pi,0,0$) and $S$ ($\pi,\pi,0$) points, the $\tilde{\mathcal{C}}_{2y}$ eigenvalues are  $\pm i$ and $\pm 1$, respectively. Since $X$ and $S$ are time-reversal-invariant momentum points, they also possess Kramers's degeneracy, which leads to the four-fold degenerate states, ($|u\rangle,\mathcal{I}|u\rangle,\Theta|u\rangle,\mathcal{I}\Theta|u\rangle$) at these points. At $X$, $\tilde{\mathcal{C}}_{2y}$ eigenvalues are purely imaginary and therefore, $|u\rangle$ and  $\Theta |u\rangle$ will have opposite sign of the screw eigenvalues. The eigenvalues of $\mathcal{I}|u\rangle,\mathcal{I}\Theta|u\rangle$ can be obtained by considering the anticommutation relation $\{\tilde{\mathcal{C}}_{2y},\mathcal{I}\}=0$ using ${\tilde{C}}_{2y}\mathcal{I} = e^{-ik_x-ik_y-ik_z}I{\tilde C}_{2y}$.
This straightforwardly leads to the eigenvalues $(+i,-i,-i,+i)$ for the operator set ($|u\rangle,\mathcal{I}|u\rangle,\Theta|u\rangle, \mathcal{I}\Theta|u\rangle$) starting with the state $|u\rangle$ with eigenvalue $+i$. Similarly, the rotational eigenvalues of the degenerate set ($|u\rangle,\mathcal{I}|u\rangle,\Theta|u\rangle,\mathcal{I}\Theta|u\rangle$)  are ($+1,+1,+1,+1$) at the $S$ point. Finally, the distinct eigenvalues at $X$ and $S$ points force a partner switching at an intermediate point along the $XS$ line, leading to an hourglass nodal point crossing on this line. This is schematically shown in Fig.~\ref{fig:CS}(b). This analysis clearly establishes that Ag$_2$BiO$_3$ supports four symmetry enforced Dirac cones in the first BZ as shown schematically in Fig. \ref{fig:wsoc}(d). To the best of our knowledge, this is the first example of a symmetry enforced 3D hourglass Dirac semimetal in an experimentally realized material.\footnote{Although our calculated hourglass Dirac cones lie in  the meV range, these predicted states could be probed via various spectroscopy techniques at low temperatures. On the other hand, the non-trivial linear energy dispersion of the hourglass states (without SOC) extends over a much larger energy range, and thus their experimental observation will not be complicated by temperature effects.}

\begin{figure}[t]
\includegraphics[width=0.5\textwidth]{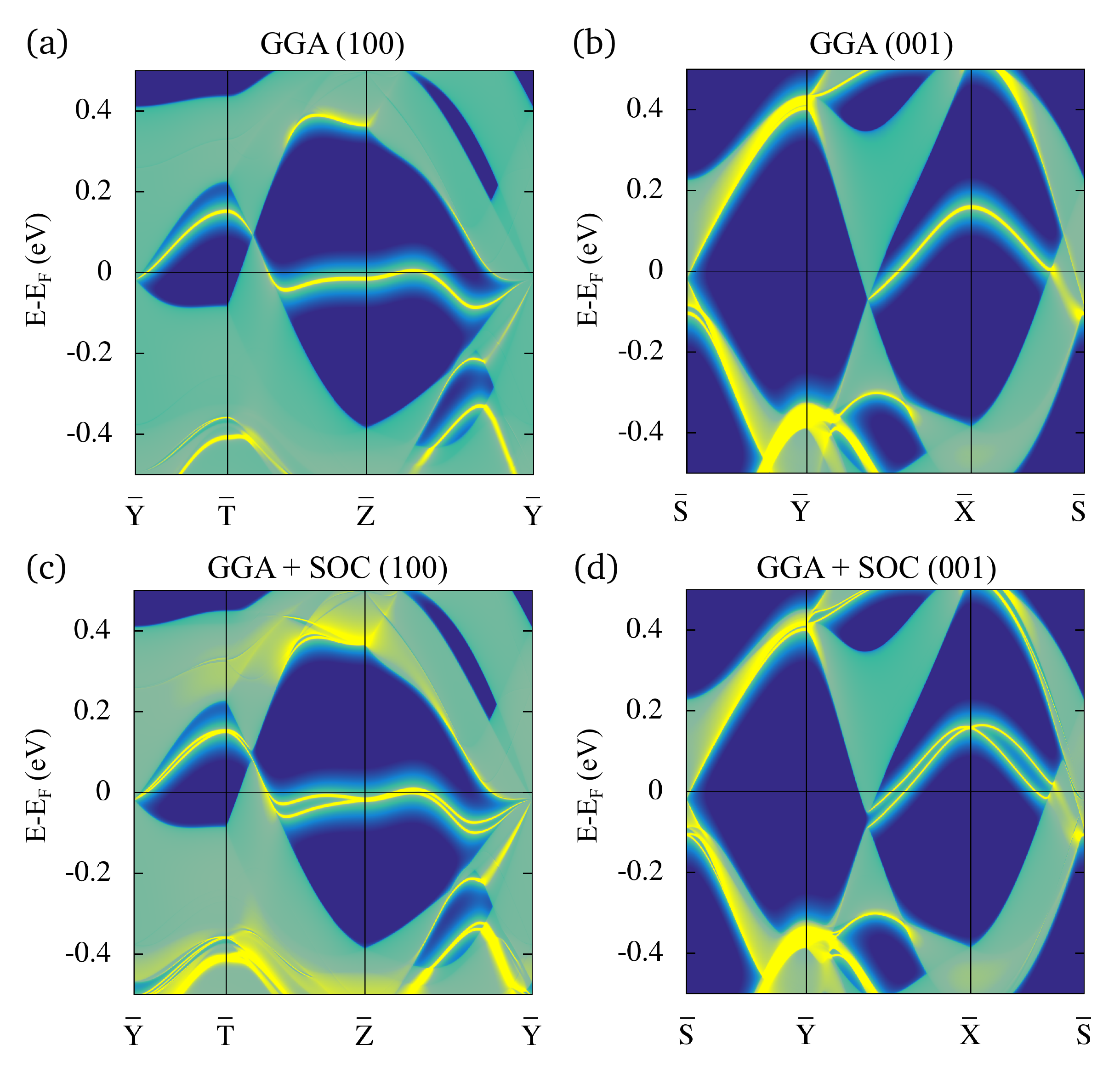}
\caption{Topological electronic structure of Ag$_2$BiO$_3$ for (a) (100) and (b) (001) surfaces without SOC. (c) and (d) are the same as (a) and (b) but with the inclusion of SOC. Sharp yellow lines identify the surface states. }  
\label{fig:SS}
\end{figure}

The nontrivial bulk band topology also manifests itself in the emergence of symmetry-protected states over various surfaces of a Ag$_2$BiO$_3$ crystal as shown in Fig.~\ref{fig:SS}. Since hourglass nodal lines (without SOC) lie on the $k_x=\pi$ and $k_z=0$ crystal planes, the nontrivial drumhead surface states would exist either inside or outside the projections of these lines on (100) and (001) crystal surfaces. The results are shown in Figs.~\ref{fig:SS}(a) and \ref{fig:SS}(b) where the drumheadlike nontrivial states are seen to reside inside the nodal-line projection for the $k_x = \pi$ plane, but lie outside the nodal-line-projection in the case of the $k_z=0$ plane. The corresponding surface states in the presence of the SOC are shown in Figs.~\ref{fig:SS}(c) and \ref{fig:SS}(d). The drumhead surface states are now gapped and deformed into Dirac conelike states. Notably, the bulk hourglass Dirac points (with SOC) project onto the $\overline{X} \overline{S}$ line close to the $\overline{S}$ point. As a result of their very small momentum separation, the surface states connecting the Dirac cones are not resolved in Fig.~\ref{fig:SS}(d).    

We emphasize that Ag$_2$BiO$_3$ crystals have been synthesized and explored experimentally \cite{DEIBELE1999117,OBERNDORFER2006267}. Neutron diffraction experiments show that Ag$_2$BiO$_3$ exists in the ferroelectric semiconducting {\it Pnn}2 phase at room temperature. However, as we have discussed above, the {\it Pnn}2 phase is a low-symmetry subgroup of {\it Pnna} that results from a Bi-O octahedral breathing distortion. Our calculations show that the energy difference between these two phases is $\sim$ 5 meV per atom in accord with earlier estimates\cite{He2018}. Since the electrical polarization is relatively large in {\it Pnn}2 Ag$_2$BiO$_3$ (8.87 $\mu$C cm$^{-2}$), the polar distortion can be suppressed by a relatively small external electric field applied in a direction opposite to that of the polarization axis\cite{He2018}. Therefore, Ag$_2$BiO$_3$ could provide an ideal testbed for exploring the ferroelectric to hourglass-semimetal transition controlled by an external electric field. 

Regarding suitable experimental probes of our predicted hourglass-semimetal state, we note that these states could be generated through electrostatic gating and probed via scanning tunneling spectroscopy (STS) experiments. Notably, pnna Ag$_2$BiO$_3$ could be grown directly using epitaxial techniques \cite{epitaxy1,epitaxy2}, which are well known for synthesizing various polymorphs of simple and complex oxides. Using the `minimal coincident interface area' and `minimum strain energy density difference' methods of substrate selection based on the Materials Project database\cite{epitaxy1}, we have identified, for example, InAs and SiO$_2$ as suitable substrates to grow this structure. 

In conclusion, based on our first-principles calculations combined with a symmetry analysis, we have shown that the nonpolar {\it Pnna} phase of Ag$_2$BiO$_3$ realizes an exotic 3D hourglass-Dirac-semimetal state. In the absence of SOC, it harbors two glide-symmetry-enforced hourglass nodal lines on the $k_x=\pi$ and $k_z=0$ planes, which are linked together to form a chainlike structure. Inclusion of the SOC gaps the nodal line crossings and results in the emergence of four symmetry-related hourglass Dirac cones in first BZ. The Dirac cones lie on the screw rotation axis and are symmetry enforced. We also show the existence of topological drumheadlike surface states over the (001) and (100) crystal surfaces. Our results show that the states on (001) surface exhibit a relatively flat electronic dispersion. Such a flat dispersion is likely to lead to many interesting correlation phenomenon and may offer a route toward realizing high $T_c$ superconductivity \cite{Sprunger1764,Liu17,singh18_saddle,JETP94_233}.

{\it Note added:---}
We recently became aware of a related preprint on Ag$_2$BiO$_3$\cite{Ag2BiO3_axriv}. Results of Ref. \cite{Ag2BiO3_axriv} are quite consistent with those presented in this study.

Work at the ShenZhen University is financially supported by the Shenzhen Peacock Plan (Grants No. 827-000113, KQTD2016053112042971), Science and Technology Planning Project of Guangdong Province (2016B050501005), and the Educational Commission of Guangdong Province (2016KSTCX126). The work at Northeastern University was supported by the US Department of Energy (DOE), Office of Science, Basic Energy Sciences Grant No. DE-FG02-07ER46352, and benefited from Northeastern University's Advanced Scientific Computation Center and the National Energy Research Scientific Computing Center through DOE Grant No. DE-AC02-05CH11231. B. G. wants to thank CSIR for Senior Research Fellowship and computer center IITK for providing HPC facility. H. L. acknowledges Academia Sinica, Taiwan for the support under Innovative Materials and Analysis Technology Exploration (AS-iMATE-107-11).  

\bibliographystyle{prsty}
\bibliography{Ag2BiO3}
\end{document}